\documentclass[prd,nofootinbib,preprint,superscriptaddress]{revtex4}
\usepackage{feynmp}
\usepackage{tikz-feynman}
\tikzfeynmanset{compat=1.1.0}
\usepackage{amsmath, amssymb, amsthm, graphicx, epsfig, fancyhdr, slashed, mathrsfs}
\usepackage{subfigure}
\usepackage{tikzsymbols}
\usepackage{natbib}
\usepackage{amsmath}
\usepackage{bbold}
\usepackage{graphicx}
\usepackage{float}

\usepackage{mathtools} 

\usepackage{tikz,xcolor,hyperref}
\hypersetup{
     colorlinks=true,
     citecolor    = green,
     linkcolor= blue
}

\definecolor{lime}{HTML}{A6CE39}
\DeclareRobustCommand{\orcidicon}{
	\begin{tikzpicture}
	\draw[lime, fill=lime] (0,0) 
	circle [radius=0.2] 
	node[white] {{\fontfamily{qag}\selectfont \tiny ID}};
	\draw[white, fill=white] (-0.0625,0.095) 
	circle [radius=0.007];
	\end{tikzpicture}
	\hspace{-2mm}
}

\foreach \x in {A, ..., Z}{\expandafter\xdef\csname orcid\x\endcsname{\noexpand\href{https://orcid.org/\csname orcidauthor\x\endcsname}
			{\noexpand\orcidicon}}
}

\newcommand{\be}{\begin{equation}}
\newcommand{\ee}{\end{equation}}
\newcommand{\bea}{\begin{equation}\begin{aligned}}
\newcommand{\eea}{\end{aligned}\end{equation}}

\allowdisplaybreaks

\begin{document}

\title{Impact of Non-Thermal Leptogenesis \\ with Early Matter Domination on Gravitational Waves \\ from First-order Phase Transition}
\author{Dilip Kumar Ghosh}
\affiliation{School of Physical Sciences, Indian Association for the Cultivation of Science, Kolkata-700032, India}

\author{Anish Ghoshal}
\affiliation{Institute of Theoretical Physics, Faculty of Physics, University of Warsaw, \\ ul. Pasteura 5, 02-093 Warsaw, Poland}

\author{Koustav Mukherjee}
\affiliation{School of Physical Sciences, Indian Association for the Cultivation of Science, Kolkata-700032, India}

\author{Nimmala Narendra}
\affiliation{Department of Physics, PES Institute of Technology $\&$ Management, Sagar Road, Shivamogga, Karnataka-577204, India}

\author{Nobuchika Okada}
\affiliation{Department of Physics, University of Alabama, Tuscaloosa, Alabama 35487, USA
}
\email{dilipghoshjal@gmail.com, anish.ghoshal@fuw.edu.pl, koustav.physics1995@gmail.com, \\ nimmalanarendra@gmail.com, okadan@ua.edu}

\begin{abstract}
\textit{
%
We study the impact of non-thermal leptogenesis on the spectrum of gravitational waves (GWs) produced 
  by a strong first-order phase transition in the early Universe.
We consider a scenario in which a heavy scalar field, $\phi$, dominates the energy density of the early Universe 
  and decays into heavy right-handed neutrinos (RHNs). 
The subsequent decay of RHNs generates a lepton asymmetry, which is partially converted into the observed baryon asymmetry 
  via the sphaleron process.
The $\phi$-dominated era and the entropy injection from the decays of $\phi$ and RHNs leave characteristic imprints 
  on the GW spectrum, such as damping and modified frequency dependence, that distinguish it 
  from the standard cosmological evolution.
We identify the parameter space in which non-thermal leptogenesis is successful, leading to distinctive GW spectral features. 
We show that these GW signals can fall within the sensitivity ranges of future detectors such as ET, DECIGO and BBO. If observed, they would provide valuable insights into the thermal history and dynamics of the early Universe.
}
\end{abstract}

\maketitle

\section{Introduction}

Evidence for physics beyond the Standard Model (SM) is confirmed by the discovery of non-vanishing tiny neutrino masses through neutrino oscillation experiments. Observations of solar \cite{dc27cfb,Super-Kamiokande:2001bfk,Super-Kamiokande:2002ujc,SNO:2002tuh,Super-Kamiokande:2005mbp,articleKam, PhysRevD.94.052010,Borexino:2015axw}, atmospheric \cite{IceCube:2017lak, ANTARES:2018rtf} and reactor \cite{KamLAND:2008dgz,T2K:2011ypd,DoubleChooz:2011ymz,T2K:2013ppw} neutrinos indicate that neutrinos are massive, and their flavor states mix due to the propagation of multiple mass eigenstates. Moreover, there are additional direct constraints on the absolute neutrino mass scale coming from the $\beta$ decay experiment KATRIN \cite{KATRIN:2021uub}; this sets an upper limit to be  $m_\nu < 0.8 $ eV. Indirect limits on the neutrino masses also arise due to cosmological observations, particularly from the measurements of the abundances of light elements via Big Bang Nucleosynthesis (BBN) \cite{Fields:2019pfx}, and the Cosmic Microwave Background radiation (CMB) by Planck 2018 \cite{Planck:2018vyg}. Large-scale structure (LSS) surveys, which measure galaxies and their distribution collectively, put an upper bound on the sum of all neutrino masses to be $\sum_i m_{\nu_i} < 0.12$ eV \cite{Planck:2018vyg,eBOSS:2020yzd}. These constraints are basically manifest due to neutrino free-streaming affecting the cosmological evolution and formation of structure.

Another compelling hint for physics beyond the SM is the observed baryon asymmetry of the Universe (BAU), which cannot be accounted for within the SM framework. The observed baryon asymmetry of the Universe is typically given in terms of the baryon to photon ratio  \cite{Planck:2018vyg}:
\begin{equation}
\eta_B = \frac{n_{B}-n_{\bar{B}}}{n_{\gamma}} = 6.1 \times 10^{-10}.
\label{etaBobs}
\end{equation}
This is also consistent with the value inferred from BBN \cite{Fields:2019pfx}.

A minimal yet compelling extension of the SM exists that can naturally accommodate both neutrino mass generation and the observed baryon asymmetry within a unified framework. If the SM is augmented via two or more Majorana right-handed neutrinos (RHNs), the well-known Type-I seesaw mechanism \cite{Minkowski:1977sc,Yanagida:1979as,Glashow:1979nm,Mohapatra:1979ia} can explain the tiny neutrino masses, while baryogenesis via leptogenesis \cite{Fukugita:1986hr} is a viable mechanism to account for generating the observed matter-antimatter asymmetry in early Universe.
In the context of thermal leptogenesis, in its minimal avatar, can elegantly explain the BAU; however, it typically requires a very high reheating temperature ($T_{\rm RH} \gtrsim 10^9$ GeV). This is closely related to what is known the Davidson-Ibarra bound, which restricts the mass of the lightest right-handed neutrino to obey $M_N \gtrsim 10^9$ GeV for a hierarchical RHN mass spectrum, and also assumes no lepton flavor effects~\cite{Nardi:2006fx,Abada:2006fw,Abada:2006ea} and an unmodified standard leptogenesis scenario~\cite{Davidson:2002qv,Giudice:2003jh}. However, such high reheating temperatures may be in tension with certain scenarios of inflationary reheating constraints. It can also lead to theoretical problems like gravitino overproduction in supersymmetric models~\cite{Kawasaki:2004qu}. Under these circumstances, alternative scenarios to thermal leptogenesis, for instance, having RHNs produced via the decay of a heavy scalar field like the inflaton in the early Universe, can be considered \cite{LAZARIDES1991305,Giudice:2000ex}. 

Despite such theoretical appeal of high-scale seesaw and leptogenesis, which accommodates neutrino mass and matter-antimatter problems in a unified framework, their direct detection and evidence remain far beyond the reach of current and foreseeable collider experiments since the RHN neutrinos need to be very heavy in the minimal version \cite{Ghoshal:2022fud}. There still can be certain indirect searches, such as evidence for lepton number violation via neutrinoless double $\beta$ decay~\cite{Cirigliano:2022oqy} and evidence for CP violation in neutrino oscillation~\cite{Endoh:2002wm}.  These considerations consequently can also impose certain theoretical constraints such as the structure of couplings consistent with ultraviolet (UV) completions like SO(10) Grand Unified Theories (GUTs) ~\cite{DiBari:2008mp,Bertuzzo:2010et,Buccella:2012kc,Altarelli:2013aqa,Fong:2014gea,Mummidi:2021anm,Patel:2022xxu} or there can also be limits from the SM Higgs vacuum (meta)stability in the early Universe ~\cite{Ipek:2018sai,Croon:2019dfw}, all of which may give us compelling insights into the viability of the whole leptogenesis mechanism to have played out in early Universe as an explanation of the observed BAU.

Having said that, let us now look at another quite promising route for testing leptogenesis via cosmological observations. With the advent of gravitational wave (GW) astronomy and GW astrophysics, the study of the GW spectrum has opened up new frontiers to test cosmology as well. The detection of GWs from black hole mergers by the LIGO and Virgo collaborations  \cite{LIGOScientific:2016aoc,LIGOScientific:2016sjg} and very recent measurements of a stochastic GW background (SGWB) from several pulsar timing array (PTA) collaborations \cite{Carilli:2004nx, Janssen:2014dka, Weltman:2018zrl, EPTA:2015qep, EPTA:2015gke, NANOGrav:2023gor, NANOGrav:2023hvm} have lead to speculations of new physics scenarios associated with GW production in the early Universe. Some of these possibilities involve primordial origins of GWs arising due to a strong cosmological first-order phase transition \cite{NANOGrav:2023hvm,Ghosh:2023aum}. Measurements of primordial GW are of great importance as they offer the intriguing possibility to probe scales of new physics much above the electroweak (EW) scale, which is beyond the reach of traditional high-energy collider experiments.

Several authors already studied the impact of the scales of seesaw and leptogenesis on cosmological observables including predictions related to Cosmic Microwave Background (CMB) spectral indices \cite{Ghoshal:2022fud}, primordial gravitational waves from local cosmic strings~\cite{Dror:2019syi, Saad:2022mzu, DiBari:2023mwu}, global cosmic strings \cite{Fu:2023nrn}, domain walls~\cite{Barman:2022yos, King:2023cgv,w8gl-wbjd}, nucleating and colliding vacuum bubbles~\cite{Dasgupta:2022isg,Borah:2022cdx}, involving hybrid topological defects~\cite{Dunsky:2021tih}, graviton bremmstrahlung~\cite{Ghoshal:2022kqp}, inflationary tensor perturbations~\cite{Berbig:2023yyy, Chianese:2025mll, Borboruah:2024eha} and light primordial blackholes~\cite{Perez-Gonzalez:2020vnz,Datta:2020bht,JyotiDas:2021shi,Barman:2021ost,Bernal:2022pue,Bhaumik:2022pil}, and dissecting the oscillatory features of the scalar curvature bi-spectrum and trispectrum involving primordial non-Gaussianities \cite{Cui:2021iie,Fong:2023egk}. In most of these scenarios described above, the primordial gravitational waves may directly probe the scale of seesaw and leptogenesis. For instance, when they are related to new physics based on new local U(1) symmetries, which is responsible for the neutrino mass paradigm, the seesaw scale is related to cosmic string tension. 

In this paper, we propose a novel scenario in which a heavy scalar field $\phi$ decays to produce an initial abundance of RHNs. The subsequent decays of these RHNs lead to non-thermal leptogenesis. Assuming the generation of gravitational waves from a strong first-order phase transition (FOPT) in the early Universe, we consider the case that a non-relativistic $\phi$ field dominates the universe's energy budget before its decay. To keep our analysis simple, we assume that this domination starts at the time of FOPT. 

We will show that on one hand, $\phi$ decay facilitates high-scale leptogenesis, while on the other hand, it also leads to a characteristic spectral shape of the GW generated by FOPT \footnote{Such epoch of early matter domination driven by $\phi$-domination were considered in general cosmology to understand the evolution history\cite{PhysRevD.31.681,Kolb:1990vq,Bezrukov:2009th}}.  Since $\phi$ decay releases a large amount of entropy, the energy density of FOPT GWs is largely diluted. We will show the capability of several upcoming GW detectors like \textit{ET, DECIGO, and BBO} to probe the parameter space for this leptogenesis scenario, which is otherwise quite elusive in conventional laboratory or astrophysical experiments.

\textit{The paper is organized as follows:} in section \ref{leptogenesis}, we describe the production of lepton asymmetry from heavy scalar decay; in section \ref{section_GW}, we discuss the GW spectral shapes from the first order phase transition; in section \ref{GW_BAU_prob} we investigate the parameter space where GW detectors probe the history of the Universe with $\phi$-domination and successful leptogenesis. Finally, we end with a conclusion and discussion in section \ref{conclusion}.

\section{Non-Thermal Leptogenesis from Heavy Scalar Decay} \label{leptogenesis}
 
We consider a heavy SM singlet scalar field $\phi$ which couples with two heavy right handed neutrinos $N_{i}$ (i=1,2) and the SM Higgs doublet H of the form\footnote{We neglect the $ \phi^2\text{H}^{\dagger} \text{H} $ term in our analysis as we are only interested in the terms that correspond to the decay of the $\phi$ field at tree level.}: 
\begin{equation}
    \mathcal{L}\supset  \lambda_{ij} \,\phi \,\overline{N^{c}_{i}} N_{j}\ + h.c.+ \lambda_{R}\,\phi \,\text{H}^{\dagger} \text{H}, 
\end{equation}
with Yukawa coupling $\lambda_{ij}$ and triple scalar coupling $\lambda_{R}$ with mass dimension 1. To simplify, we assume $2M_{N_1}<M_{\phi}<M_{N_2}$, so that the $\phi$ can decay only to $N_1 N_1$ and 
$\text{H}^{\dagger} \text{H}$, where $M_{N_i}$ and $M_{\phi}$ are the mass of the right-handed neutrinos and the scalar field $\phi$, respectively. We consider the scenario that $\phi$ decays when its energy density dominates over the pre-existing radiation energy density and reheats the Universe by the decay products, $N$'s, and H.  

The subsequent decay of the heavy right-handed neutrinos generates lepton asymmetry, which is then transferred to the baryon asymmetry of the Universe via the sphalerons. Here, we assume that $T_{RH}< M_N$ for the reheating temperature $T_{RH}$, so that the right-hand neutrino remains out of equilibrium. In this non-thermal leptogenesis scenario, the wash-out processes are all inactive, and the resultant lepton asymmetry can be easily estimated. For simplicity, we set the lifetime of the produced right handed neutrino ($\tau_{N}$) to be much shorter than the lifetime of $\phi$ ($\tau_{\phi}$), i.e., $\tau_{N} \ll \tau_{\phi}$, so that the right handed neutrinos, once produced, immediately decay.
 
The relevant Lagrangian for the RHNs can be expressed as \cite{Minkowski:1977sc,Yanagida:1979as,Glashow:1979nm,Mohapatra:1979ia}
\begin{eqnarray}
   - \mathcal{L} & \supset & Y_{i\alpha} \,\overline{N_{i}} \text{H}^{\dagger} L_{\alpha} + (M)_{ii} \overline{N_i^{c}} N_i + h.c. ,
   \label{Lagrangian}
\end{eqnarray}
where $Y_{i\alpha}$ is the Yukawa coupling, with $\{i=1,2\}$ and $\{\alpha=1,2,3\}$, $L_{\alpha}$ is the lepton doublet, and H is the Higgs doublet. Without loss of generality, we work in the basis where the Majorana mass matrix of RHNs, which violates the lepton number, is diagonal. The spontaneous electroweak symmetry breaking by the Higgs vacuum expectation value (VEV) generates the neutrino Dirac mass term, and via the Type-I Seesaw mechanism, the light neutrino mass matrix can be expressed as
\begin{equation}
    m_\nu \simeq m_{D} M^{-1} m_{D}^T,
\end{equation}
where $m_{D}= Y v/\sqrt{2}$, with $v=246$ GeV being the SM Higgs VEV. We can diagonalize this matrix by a unitary transformation,
\begin{equation}
    D_\nu = \text{diag}(m_1,m_2,m_3)= U^T m_\nu U ,
\end{equation}
where $U$ is the Pontecorvo-Maki-Nakagawa-Sakata (PMNS) matrix. We work on the flavor basis where the charged-lepton mass matrix is diagonal. 

We can express the Yukawa coupling matrix in terms of the Casas-Ibarra parametrization as \cite{Casas:2001sr} 
\begin{equation}
Y = \frac{\sqrt{2}}{v} U \sqrt{D_\nu} R^T \sqrt{M} ,
\label{CI_param}
\end{equation}
where $R$ is a $2\times 3$ complex matrix satisfying $R R^T=\mathbb{1}$ and $\sqrt{M}=\text{diag}(\sqrt{M_{N_1}}, \sqrt{M_{N_2}})$.
The dynamics of Leptogenesis are governed by the following entity:
\begin{equation}
Y^{\dagger}Y = \frac{2}{v^2} \sqrt{M} R^{*}D_{\nu} R^T \sqrt{M} .
\label{CI_param_yy}
\end{equation}
 Note that Eq.\,(\ref{CI_param_yy}) is independent of the PMNS matrix $U$. In our scenario, we consider both Normal Hierarchy (NH) where $m_{1}<m_{2}<m_{3}$ and Inverted Hierarchy (IH) where $m_{3}<m_{1}<m_{2}$, for the light neutrino masses.
 For NH case, the $R$ and $D_{\nu}$ matrices are given by,
\begin{equation}
R=\begin{pmatrix}
  0&\cos z' & \sin z'  \\
  0&-\sin z' & \cos z'  \\
\end{pmatrix} ~~~~~~~~ , ~~~~~~~~~ \sqrt{D_{\nu}}=\text{diag}(0, \sqrt{m_2}, \sqrt{m_3})
\end{equation}
while for the IH case, 
\begin{equation}
R=\begin{pmatrix}
  \cos z' & \sin z' & 0 \\
 -\sin z' & \cos z' & 0 \\
\end{pmatrix} ~~~~~~~~~, ~~~~~~~~~ \sqrt{D_{\nu}}=\text{diag}(\sqrt{m_1}, \sqrt{m_2}, 0).
\end{equation}
where $z'=a' + i \, b'$ is a complex angle, with two real parameters, $a'$ and $b'$. 
Note that in the Type-I Seesaw model with two RHNs, the lightest of the three neutrinos is massless. By using the neutrino oscillation data for the mass-squared differences, $\Delta m_{21}^2$ (solar) and $\Delta m_{32}^2$ (atmospheric), which are given by 
\begin{eqnarray}
    \Delta m_{21}^2 \equiv m_2^2 - m_1^2,~~~
\Delta m_{32}^2 \equiv m_3^2 - m_2^2,
\end{eqnarray}
and the mass eigenvalues for NH are determined as,
\begin{eqnarray}
    m_2 = \sqrt{\Delta m_{21}^2}, \quad m_3 = \sqrt{\Delta m_{32}^2 + \Delta m_{21}^2},
\end{eqnarray}
where \( \Delta m_{21}^2 = 7.53 \times 10^{-5}~\text{eV}^2 \) and 
\( \Delta m_{32}^2 = 2.455 \times 10^{-3}~\text{eV}^2 \)~\cite{ParticleDataGroup:2024cfk}. 
While for the IH case, the masses are
\begin{eqnarray}
    m_1 = \sqrt{-\Delta m_{32}^2 - \Delta m_{21}^2}, \quad m_2 = \sqrt{-\Delta m_{32}^2},
\end{eqnarray}
where \( \Delta m_{21}^2 = 7.53 \times 10^{-5}~\text{eV}^2 \) and \( \Delta m_{32}^2 = -2.529 \times 10^{-3}~\text{eV}^2 \)~\cite{ParticleDataGroup:2024cfk}.

\subsection{The Boltzmann Equations}
The relevant Boltzmann equations governing the evolution of the energy densities of the scalar field $\phi$ ($\rho_{\phi}$), radiation ($\rho_{R}$), the lightest of the right-handed neutrinos $N_1$ ($\rho_{N}$) and the $B-L$ asymmetry which is created due to decay of $N_1(\equiv N)$ are given by
\begin{equation}
\begin{array}{rl}
\displaystyle \frac{d \rho_\phi}{dt}+3 H \rho_\phi &= -\Gamma_{\phi} \,\rho_\phi, \\[1ex]
\displaystyle \frac{d \rho_R}{dt}+4 H \rho_R &= B_R\, \Gamma_{\phi} \,\rho_\phi + \Gamma_{N} \rho_N, \\[1ex]
\displaystyle \frac{d \rho_N}{dt}+3 H \rho_N &= B_{NN}\, \Gamma_{\phi}\, \rho_\phi - \Gamma_{N} \rho_N, \\[1ex]
\displaystyle \frac{d n_{B-L}}{dt}+3 H n_{B-L} &= \frac{\epsilon_1}{M_{N}} \Gamma_{N} \rho_N.
\end{array}
\label{BEq_1}
\end{equation}
Here, $\Gamma_\phi$ is the total decay width of $\phi$ and is given by
\begin{eqnarray}
    \Gamma_\phi &\equiv & \Gamma_{\phi}^{NN}+\Gamma_\phi^R \, \\
    & = & (B_{NN}+B_{R}) \Gamma_\phi  \,\,,
\end{eqnarray}
where $\Gamma_{\phi}^{NN}$ and $\Gamma_{\phi}^{R}$ are the partial decay widths of the scalar field $(\phi)$ to a pair of right-handed neutrinos $\phi \rightarrow N + N $ and a pair of Higgs doublets , and $B_{NN}=\Gamma_{\phi}^{NN}/\Gamma_{\phi}$ and $B_{R}=\Gamma_{\phi}^{R}/\Gamma_{\phi}$ are branching ratios. 
These partial decay widths are calculated to be
\begin{eqnarray}
    \Gamma_\phi^{NN} &=& \frac{|\lambda|^2}{4 \pi} M_{\phi} \left(1-\frac{4 M_{N}^{2}}{M_\phi^{2}} \right)^{\frac{3}{2}} ~~~,~~~~
    \Gamma_{\phi}^{R}=\frac{|\lambda_R|^{2}}{4 \pi M_{\phi}},
    \label{Gamma_phi_decay_widths}
\end{eqnarray}
 where $\lambda\equiv\lambda_{11}$.
 
The $\Gamma_{N}$ is the decay width of the right-handed neutrino to SM particles, $N \rightarrow L + \text{H}$, and is given by, 
\begin{equation}
    \Gamma_N = \frac{(m_D^{\dagger}m_D)_{11}}{4\pi v^2} M_{N}.
\end{equation} 
Note that we neglect the lepton asymmetry washout processes since we set $T_{RH}<M_N$ and the thermal bath lacks the required energy to reproduce right-handed neutrinos, where $T_{\text{RH}}$ is the reheating temperature of radiation created by $\phi$ decay and is estimated by
\begin{equation}
    T_{\text{RH}}=\left( \frac{90}{8\pi^{3}g_*} \right)^{1/4} \sqrt{\Gamma_\phi M_{\text{Pl}}},
\end{equation}
where $M_{Pl}~(1.2\times10^{19} \rm{GeV})$ is the Planck mass.

The existence of CP violation in this framework results in the emergence of the CP asymmetry parameter $\epsilon_1$ in the last line of Eq. (\ref{BEq_1}), which is defined as
\begin{equation}
    \epsilon_{1} \equiv \frac{\Gamma(N_1\rightarrow L\text{H})-\Gamma(N_1\rightarrow \bar{L}\text{H}^{\dagger})}{\Gamma(N_1\rightarrow L\text{H})+\Gamma(N_1\rightarrow \bar{L}\text{H}^{\dagger})}.
\end{equation}
This CP-asymmetric parameter is calculated by an interference between the tree-level and one-loop level processes of $N_{1}$ decay:
\begin{eqnarray}
\label{cpasym}
        \epsilon_1= \frac{3}{16 \pi (Y^{\dagger}Y)_{ii}} \sum_{j\neq i} \text{Im}\left[(Y^{\dagger}Y)_{ji}^{2}\right] \left( \frac{M_{N_i}}{M_{N_j}} \right).
        \label{epsilon1}
\end{eqnarray} 
From Eq.\,(\ref{CI_param_yy}), the CP-asymmetric parameter can be expressed as follows:
For the NH case,
\begin{equation}
    \epsilon_{1}^{\text{NH}}=\frac{3 M_{N_1}}{16 \pi v^2} \frac{\Delta m_{32}^2 \,\text{sin}(2a')\, \text{sinh}(2b')}{\{ (m_2-m_3) \,\text{cos}(2a') + (m_2+m_3) \,\text{cosh}(2b') \}}\,
\label{CPNH}
\end{equation}
For the IH case,
\begin{equation}
    \epsilon_{1}^{\text{IH}}=\frac{3 M_{N_1}}{16 \pi v^2} \frac{\Delta m_{21}^2 \,\text{sin}(2a') \,\text{sinh}(2b')}{\{ (m_1-m_2) \,\text{cos}(2a') + (m_1+m_2) \,\text{cosh}(2b') \}}\,.
    \label{CPIH}
\end{equation}
From Eq.(\ref{CPNH}) and Eq.(\ref{CPIH}), the CP-asymmetry parameter depends on the mass squared differences and the real and imaginary parts of the complex angle $z'$, for a fixed value of mass scale $M_{N_1}$.
For convenience, we use the following quantities for the numerical evaluation of the Boltzmann equations given in Eq.(\ref{BEq_1}) :
\begin{eqnarray}
E_{\phi}&=&\rho_{\phi} a^{3}, ~~~~~R=\rho_{R} a^{4},\nonumber\\
E_{N}&=&\rho_{N} a^{3}, ~~~~~~\widetilde{N}_{B-L}=n_{B-L} a^{3}
\label{ini_cond}
\end{eqnarray}
where $a$ is the scale factor of the Universe. 

For convenience, we write the Boltzmann equations as functions of the scale factor rather than time. We define a variable $y$ as the ratio of the scale factor to its initial value,
\begin{equation}
y=\frac{a}{a_I}  ~,
\end{equation}
which acts as a time variable. We assume the initial value of the scale factor to be $a_I=1$ since no physical result depends on this choice \cite{Giudice:2000ex, Barman:2021ost}. With these definitions, the Hubble parameter reads
\begin{equation}
    H=\sqrt{\frac{8\pi}{3 M_{Pl}^{2}} \frac{(a_I E_\phi y +a_I E_N y +R)}{a_I^{4} y^{4}} }.
\end{equation}
Further, we define a dimensionless variable $z=M_{N}/T$ and express it in terms of $R$ as
\begin{equation}
    z=\frac{M_{N}}{T}=M_{N} \left[\frac{\pi^{2} g_{*}}{30 \, R} \right]^{1/4} a_I \, y\,.
\end{equation}
In terms of these newly defined rescaled quantities, the set of Boltzmann equations can be re-expressed as
\begin{eqnarray}
    \frac{d E_\phi}{dy} &=& -\frac{\Gamma_{\phi}}{H y} E_\phi, \nonumber\\
    \frac{d R}{dy} &=& B_{R}\frac{\Gamma_{\phi}}{H} a_I E_\phi + \frac{\Gamma_{N}}{H} a_I E_N, \nonumber\\
    \frac{d E_N}{dy} &=& B_{NN}\frac{\Gamma_{\phi}}{H y} E_\phi-\frac{\Gamma_{N}}{H y} E_N, \nonumber\\
    \frac{d \widetilde{N}_{B-L}}{dy} &=& \frac{\epsilon_1}{M_{N}} \frac{\Gamma_{N}}{H y} E_N.
    \label{BEq_2}
\end{eqnarray}

At early times, the energy density of the Universe was dominated by the heavy scalar field, $\phi$. We set the initial density of $\phi$ as \cite{Hahn-Woernle:2008tsk}:
\begin{equation}
    \rho_{\phi_I} =\frac{3 M_{Pl}^{2}}{8\pi} M_\phi^2.
\end{equation}
To solve the set of Boltzmann equations given in Eq.(\ref{BEq_2}) we choose the initial conditions at $a=a_I$ as: $R(a_I)=0, E_{N}(a_I)=0, \widetilde{N}_{B-L}(a_I)=0$ and $E_{\phi}(a_I)\equiv E_{\phi_I}=(3/8\pi) M_{Pl}^{2} M_\phi^2 a_I^3$ (see, Eq.(\ref{ini_cond})). Note that the numerical value of $a_I$ is irrelevant. 

The $\widetilde{N}_{B-L}$ asymmetry is related to the total $N_{B-L}$ via \cite{Hahn-Woernle:2008tsk},
\begin{equation}
    N_{B-L}=\frac{n_{B-L}}{n_{\gamma}}=\left(\frac{\pi^4}{30 \,\zeta(3)}\right) \left(\frac{30}{\pi^{2}}\right)^{1/4} \frac{g_{*}^{3/4}}{g_{\gamma}} R^{-3/4} \widetilde{N}_{B-L},
    \label{N_BL}
\end{equation}
 where we re-expressed $n_\gamma=\frac{2 \zeta(3)}{\pi^{2}} T^3$ in terms of $\rho_\gamma=\frac{2 \pi^2}{30} T^{4}$, and $T$ in terms of $\rho_R$, where $\rho_R=\frac{\pi^2 g_{*}}{30} T^{4}$.
The generated $N_{B-L}$ asymmetry obtained from Eq.(\ref{N_BL}), converts into the baryon asymmetry via the sphaleron processes. The predicted $N_{B-L}$ is related to the $\eta_B$ measured at recombination given as
\begin{equation}
    \eta_B= \left(\frac{a_{sph}}{f} \right)  N_{B-L} \,,
    \label{etaB}
\end{equation}
where $a_{sph}=28/79$ is the fraction of $B-L$ asymmetry converted into a baryon asymmetry by sphaleron processes, and $f=N_\gamma^{\text{rec}}/N_\gamma^{\text{*}}=2387/86$ is the dilution factor calculated assuming standard photon production from the onset of Leptogenesis till recombination \cite{Buchmuller:2004nz}.

From Planck data \cite{Planck:2018vyg}, $\eta_{B}$ is given by Eq.(\ref{etaBobs}). At the present era, the entropy density for the relativistic degrees of freedom $s$ is expressed as $s = 7.04 n_\gamma $, where $n_\gamma$ is 
the photon number density. Therefore, $Y_B$ can be expressed as:
\begin{equation}
    Y_B=\frac{n_B}{s}=8.7\times 10^{-11}\,.
\end{equation}

\begin{figure}[ht]
    \centering
	\includegraphics[height = 7cm,width=10cm]{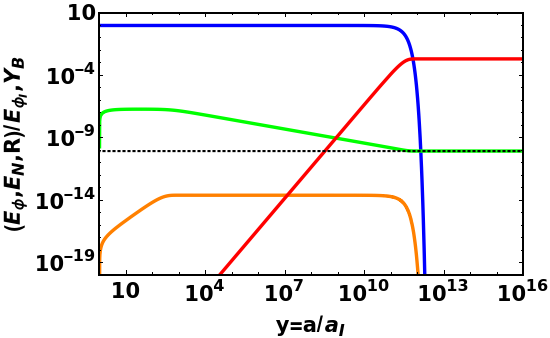}
	\caption{\it  The abundances of $\phi$(Blue), RHN (Orange), $Y_{B}$ (Green) and radiation (Red). The Black dotted line is the observed baryon asymmetry of the Universe. The parameter choices are as follows: $M_{\phi}=10^{14}$ GeV, $M_{N}=10^{13}$ GeV, $\Gamma_{\phi}^{NN}=4.7\times10^{-3}$ GeV, $\Gamma_{\phi}^{R}=7.95\times10^{-16}$ GeV, $\epsilon_{1}=1.6\times10^{-4}$. }
	\label{BAU_y}
    
\end{figure}

We numerically solve the set of Boltzmann equations given in Eq.(\ref{BEq_2}) and demonstrate the variation of $\{E_{\phi},\, E_N,\, R,\,\text{and},\,Y_B \}$ as a function of $y(=a/a_I)$ in Fig.\ref{BAU_y} for a sample parameter set listed in the caption. We normalize the $\{E_{\phi},\, E_N,\, R \}$ yields with the initial scalar field energy density $E_{\phi_I}$, such that they turn out to be dimensionless quantities. We show the yields of the quantities $\{ \,\phi,\, N,\, R,\, \text{and}\,Y_B \}$ with Blue, Orange, Red, and Green lines, respectively. The scalar field dominates the energy density of the Universe before the Hubble rate falls below the decay rate of the scalar field, $\Gamma_{\phi}>H$. As the Hubble rate falls below the decay rate, $\phi$ quickly decays away completely. We assume the branching ratio $B_{NN}$ is much larger than the branching ratio of radiation, $B_{NN}\gg B_{R}$, since $\Gamma_{\phi}^R$ is suppressed by $M_{\phi}$. The produced RHN decay width is much larger than the total decay width of $\phi$ {\it i.e.,} $\Gamma_{N} \gg \Gamma_{\phi}$. The energy budget of the Universe $\rho_{\phi}$ transfer to $\rho_{N}$, at time $t \sim 1/\Gamma_{\phi}$. The produced RHNs decay instantaneously, once created by the $\phi$ decay, and the radiation is created, {\it i.e., $\rho_\phi \sim \rho_R$}. We consider that the $\phi$ decay reheats the Universe to a temperature $T_{RH} < M_N$.

In Fig.\ref{BAU_y}, the production and the decay of the right-handed neutrinos compete and cause a plateau region in the yield of right-handed neutrinos. When the scalar field decay is completed, the right-handed neutrino production stops, and it quickly decays away. With the scalar decay being the source of right-handed neutrino production, the yields of $\phi$ (Blue line) and $N$ (Orange line) fall at the same point. Once the heavy right-handed neutrino is produced from the scalar decay, it subsequently decays, producing the $N_{B-L}$ ($\eta_B$ from Eq. (\ref{etaB})) asymmetry and the radiation $R$. Initially, the $\widetilde{N}_{B-L}$ and $R$ increase together linearly, which cause a plateau region for $N_{B-L}$ (${\it i.e.,} \eta_B$ or the $Y_B$, the Green line), as a result of Eq.\,(\ref{N_BL}). The $N_{B-L}(Y_B)$ plateau region continues until the right-handed neutrino production and its decay start to compete with each other ($N$ plateau region). In the $N$ plateau region, the production rate of $N_{B-L}$ decreases compared to radiation, which causes a small inclination (radiation suppresses dominantly, see Eq.\,(\ref{N_BL})) of $N_{B-L}$ to occur. This inclination continues until the right-handed neutrino completely decays away. When the right-handed neutrino completely decays away, the $N_{B-L} (Y_B)$ asymmetry becomes a constant.

One can also analytically approximate the Boltzmann Equations in Eq.(\ref{BEq_1}) and estimate the final baryon asymmetry. From the last equation of Eq.(\ref{BEq_1}), we can see that $n_{L}=\epsilon~n_{N}$. Finally, the lepton asymmetry will be proportional to the baryon asymmetry, $n_{B}$. Thus, we estimate the quantity $\frac{n_{B}}{s}$ as
\begin{equation}
 Y_B\simeq\frac{n_{B}}{s}=\epsilon_1
 \left(\frac{\rho_{\phi}}{M_{N}}\right)/\left(\frac{\rho_{\phi}}{T_{RH}}\right)
 =\epsilon_1
 \left(\frac{T_{RH}}{M_{N}}\right).
 \label{yb_analy}
\end{equation}
We can check that this rough estimate is consistent with the numerical solution of the Boltzmann equations.

\section{Gravitational Waves from FOPT}
\label{section_GW}
\subsection{GW spectral shape analysis}

We consider the presence of stochastic gravitational waves arising from a cosmological first-order phase transition in the early Universe. Generally, the profile of the amplitude of the gravitational wave spectrum depends on the underlying model parameters.
We analyze the spectral properties of gravitational waves by using a model-independent approach, allowing us a thorough examination that is free from specific theoretical constraints. This ensures our findings reflect the intrinsic characteristics of the gravitational wave spectrum, offering insights across various cosmological and astrophysical contexts.

The cosmological first-order phase transitions in the early Universe are associated with the evolution of the Universe from a metastable false vacuum to a true vacuum. We analyse such a phase transition based on the effective potential at finite temperature $T$. 
The effective potential at the one-loop level can be expressed as \begin{equation}
     V_{\rm eff}(\Phi_{i},T)=V_{\rm tree}(\Phi_{i})+ V_{\rm CW}^{\rm 1-loop}+ V_{\rm ct}+V_{T\neq 0}^{\rm 1-loop},\label{eq:totPot}
\end{equation}
where $\Phi_{i}$ is any scalar field of the theory under consideration, $V_{\rm tree}(\Phi_{i})$ is the tree level potential, $V_{\rm CW}^{\rm 1-loop}$ and $V_{\rm ct}+V_{T\neq 0}^{\rm 1-loop}$ are the Coleman-Weinberg potential at zero and finite temperature, respectively, and $ V_{\rm ct}$ is the counter term at zero temperature which ensures the masses and VEVs to retain their zero temperature values. Our primary interest is to find the relevant
parameters from the effective potential that control the spectral properties of gravitational waves produced by the cosmological first-order phase transition.

The GW spectrum depends mainly on four parameters, namely $\alpha$, $\frac{\beta}{H_{*}}$, $T_{*}$, $v_{w}$:\\1) $T_{*}$ represents the nucleation temperature associated with the phase transition, \\2) $\alpha$ signifies the strength of the phase transition, which is proportional to the latent heat produced during the phase transition and is expressed as \begin{equation}
    \alpha = \frac{\epsilon(T_{*})}{\rho_{R}(T_{*})}\,,
\end{equation}
where $\epsilon$ is the latent heat generated during the transition and is expressed as\begin{equation}
    \epsilon(T_{*})= \Delta V_{eff} - T\frac{d\Delta V_{eff}}{dT}\bigg|_{T=T_{*}}\,.
\end{equation}
    Here, $\Delta V_{eff}$ is the difference between the effective potentials at false and true vacua, and $\rho_{R}(T_{*})$ is the energy density of radiation given by
    \begin{equation}
        \rho_{R}(T_{*})=\frac{\pi^2 g_{*}T_{*}^{4}}{30}
    \end{equation}
    with $g_{*}$ representing the relativistic degrees of freedom at $T_{*}$ ,
\\3) $\frac{\beta}{H_{*}}$ denotes the ratio of the inverse of the time taken for the phase transition to complete to the Hubble parameter value at $T_{*}$ and can be expressed as
\begin{equation}
    \frac{\beta}{H_{*}}= T_{*}\frac{d(S_{3}/T)}{dT}\bigg|_{T=T_{*}}\,,
\end{equation}
where $S_{3}$ represents the 3-dimensional Euclidean Action,
\begin{equation}
    S_{3}=4\pi \int  r^2 dr\left[\frac{1}{2}\left(\frac{d\Phi}{dr}\right)^2 + V_{eff}(\Phi,T)\right]\,;
\end{equation}\\4) $v_{w}$ represents the velocity of the bubble walls when bubbles of the true vacuum collide with each other and the energy is released in the form of gravitational waves. In our analysis, this velocity is taken to be 1 (relativistic) as we are considering supercooled phase transitions which have a high value of $\alpha$. 
As we are only interested in the shape of the GW spectrum, we present a model-independent analysis by fixing the key parameters: $\alpha$, $\beta/H_*$, and $T_{*}$. For any model with an arbitrary number of scalar fields, suitable choices of model parameters can reproduce these fixed values, thereby making our analysis applicable within the context of such models.

The main processes that are responsible for the generation of Gravitational Waves from first-order phase transitions are:\\1) Bubble Collision: During the transition, bubbles of the true vacuum nucleate and fill the entire space. When these bubbles collide, the energy of the collision sources gravitational waves. In our work, we use the envelope approximation \cite{PhysRevD.47.4372}, which assumes that GW production is dominated by the uncollided portions of the bubble walls and that the contribution from overlapping regions vanishes instantaneously upon collision. \\2) Sound waves generated in the plasma: As the bubbles expand, they deposit energy into the surrounding plasma, generating bulk fluid motions. These motions cause the sound waves that source the gravitational waves \cite{Hindmarsh:2013xza}.\\3) Magnetohydrodynamic (MHD) turbulence in the plasma: The acoustic waves thus generated in the plasma can create turbulent flows and sometimes get coupled to primordial magnetic fields, thus creating MHD Turbulence. This turbulence sources gravitational waves \cite{Caprini:2009yp}. 
The total Gravitational Wave amplitude spectrum is thus given by 

\begin{equation}
    \label{3.1}
    \Omega_{GW} h^2 \simeq \Omega_{col} h^2+ \Omega_{sw} h^2+ \Omega_{turb} h^2\,.
\end{equation}
\subsection*{Bubble Collisions}

%

 The GW spectrum as a function of frequency generated by bubble collisions during the phase transition at temperature $T_*$ is given by \cite{Ellis:2020nnr}

\begin{equation}
\label{omgcol}
\resizebox{\textwidth}{!}{$
\Omega_{col,*} h^2(f) = 2.3 \times 10^3 \left( \frac{R_* H_*}{\sqrt[3]{8\pi}} \right)^2 \left( \frac{\kappa_{col} \alpha}{1 + \alpha} \right)^2 \left[ 1 + \left( \frac{f}{f_{d,*}} \right)^{-1.61} \right] \left( \frac{f}{f_{col,*}} \right)^{2.54} \left[ 1 + 1.13 \left( \frac{f}{f_{col,*}} \right)^{2.08} \right]^{-2.3}
$},
\end{equation}
where $R_{*}$ is the radius of the bubble wall, and $\kappa_{col}$ is the efficiency factor, which is proportional to the fraction of energy released due to the bubble collision in the GW spectrum.
 The frequency $f_{d,_{*}}$ is related to $R_{*}$ by \begin{equation}
     f_{d,*} \simeq \left(\frac{\ 0.044}{R_{*}}\right),
 \end{equation} and the peak frequency $f_{col,*}$ of collisions
 \begin{equation}
     f_{col,*} \simeq \left(\frac{\ 0.28}{R_{*}}\right).
 \end{equation}

\subsection*{Sound Waves}
The GW spectrum generated by the sound waves in the plasma is given by \cite{Ellis:2020nnr,Banik:2025olw} 
\begin{equation}
\label{omgsw}
    \Omega_{sw,*} h^2(f)=0.384\left(R_* H_* \right) \left(\tau_{sw} H_*\right) \left(\frac{\kappa_{\text{sw}} \alpha}{1+\alpha} \right)^{2} \left( \frac{f}{f_{sw,*}} \right)^{3} \left[1+\frac{3}{4}\left(\frac{f}{f_{sw,*}}\right)^{2} \right]^{-7/2}\,,
\end{equation}
where the peak frequency of the sound waves at $T_{*}$ is given by
\begin{equation}
    f_{sw,*} \simeq \left(\frac{3.4}{R_{*}}\right),
\end{equation}
The duration of the sound wave period is given by 
\begin{equation}
    \tau_{sw} =\frac{R_* }{U_{f}}~,~~~~ U_{f} \simeq \sqrt{\frac{3}{4} \frac{\alpha}{1+\alpha} \kappa_{sw}},
\end{equation}
where $\kappa_{sw}$ is the efficiency factor which is proportional
to the fraction of energy released due to the propagation of sound waves in the plasma.

\subsection*{Magnetohydrodynamic Turbulence}

The part of the GW amplitude spectrum generated due to the MHD Turbulence generated in the plasma is given by \cite{Ellis:2020nnr} 

\begin{equation}
\Omega_{\text{turb},*}(f)\, h^2 = 6.85 \left( \frac{R_* H_*}{1 - \tau_{\text{sw}} H_*} \right) 
\left( \frac{\kappa_{\text{sw}} \, \alpha}{1 + \alpha} \right)^{3/2}
\left( \frac{f}{f_{\text{turb}}} \right)^3
\left[ 1 + \left( \frac{f}{f_{\text{turb}}} \right) \right]^{-11/3}
\left( 1 + \frac{8\pi f}{H_*} \right)^{-1},
\end{equation}
where the peak frequency of the MHD Turbulence at $T_{*}$ is given by
\begin{equation}
    f_{turb,*} \simeq \left(\frac{5.1}{R_{*}}\right)\,.
\end{equation}

\subsection*{Redshifting of GW to today}
As we know, after their generation, the stochastic GWs from the first-order phase transition redshift to the present time. Since graviton is massless, GWs redshift in the same way as that of radiation \cite{Ellis:2020nnr}. 
As expected, the early matter domination (EMD) phase alters the GW spectrum 
compared to the standard radiation domination scenario. The red-shifting of the 
GW in the presence of the EMD is described as:
\begin{equation}
\label{GWred}
\Omega_{\text{GW,0}}(f) =
\begin{cases} 
\left( \frac{a_*}{a_0} \right)^4 \left( \frac{H_*}{H_0} \right)^2 \Omega_{\text{GW,*}} \left( \frac{a_0}{a_*} f \right) & \text{for } f > f_*, \\[10pt]
\left( \frac{a_f}{a_0} \right)^4 \left( \frac{H_f}{H_0} \right)^2 \Omega_{\text{GW,*}} \left( \frac{a_0}{a_*} f_* \right) \left( \frac{f}{f_*} \right)^3 & \text{for } f < f_*,
\end{cases}
\end{equation}
where $a_{f}$ denotes the scale factor when the scale $2\pi f$ re-enters the Hubble horizon in the expanding Universe, {\it i.e.,} when $a_{f} H_{f}=2\pi f$. We use the symbol $a_{*}$ to denote the scale factor at the time of phase transition.

\subsection{Imprints of $\phi$-domination on the GW signal} 

Our primary objective lies in investigating the manifestations of an early matter domination effect driven by the scalar $\phi$. This effect will leave imprints in the GW spectrum during the early stages of the Universe. After its domination, the scalar $\phi$ field decays into right-handed neutrinos and radiation. As we will see, the decay width of the scalar field plays an important role in shaping the GW spectrum while it also controls the production of RHNs and its abundance, thus we allude to a correlation between the final GW spectral shape at present and baryogenesis via leptogenesis, which is otherwise challenging to test in collider physics due to the energy scales being too high. The intermediate $\phi$-dominated era leads to a modification of the cosmic expansion history of the Universe, leaving characteristic imprints on the shapes of the GW spectrum via the redshift factor and a modification of the GW amplitude due to entropy injection from $\phi$ decay.

Let us denote the Hubble parameter during this period (from $t_{*}$ to $t_{reh}$) as a function of scale factor to be \cite{Ellis:2020nnr}
\begin{equation}
    \label{Hexp}H(a)=H_{*}\left(\frac{a_{*}}{a_{reh}}\right)^{\frac{3}{2}}\left(\frac{a_{reh}}{a}\right)^{2}\left[1+\left(\frac{a_{reh}}{a}\right)^{2}\right]^{-\frac{1}{4}}.
\end{equation}
It is important to note that the ratio of the scale factors at the time of this secondary reheating caused by $\phi$ decay, and at the time of phase transition, ($\frac{a_{reh}}{a_{*}}$) has to be significantly greater than unity to ensure $\Gamma_{\phi} \ll H_{*}$ is satisfied in the matter-dominated era.

Let us look at the expression for the lifetime of $\phi$ ($\tau_{\phi}$), which is the inverse of the decay width given by
\begin{equation}
    t_{reh}-t_{*}=\tau_{\phi}=\frac{1}{\Gamma_{\phi}}=\int_{a_*}^{a_{reh}}\frac{da}{a\,H}= \frac{2}{3}\frac{1}{H_{*}}\left[\left(\frac{a_{reh}}{a_*}\right)^{\frac{3}{2}}-1\right].
    \label{dur_MD}
\end{equation}
Next we obtain the ratio of the scale factors for $\Gamma_{\phi} << H_{*}$ to be,
\begin{equation}
\label{arehas}
    \frac{a_{reh}}{a_{*}}\simeq\left(\frac{3H_{*}}{2\Gamma_{\phi}}\right)^\frac{2}{3} \gg 1.
\end{equation}
Utilising Eq.(\ref{Hexp}) and Eq. (\ref{arehas}), the evolution of the scale factor during the EMD epoch is given by
\begin{equation}
\label{aareh}
    \frac{a}{a_{\rm reh}} = \sqrt{\frac{\xi(f)}{2} - \frac{1}{2}},
\end{equation}
where $\xi(f)$ is basically
\begin{equation*}
    \xi(f) = \sqrt{8 \left(\frac{f_{H_{reh}}}{f}\right)^{4}+1},
\end{equation*}
     and $f_{H_{reh}}$ is the frequency when the scale factor ($a_{reh}$) corresponds to the time of reheating.
Using Eq.(\ref{aareh}) in the GW red shifting spectrum, Eq. (\ref{GWred}) one has the corresponding redshift until $a_{reh}$ to be
\begin{equation}
\label{finred}
\left(\frac{a}{a_{\rm reh}}\right)^{4} \left(\frac{H}{H_{\rm reh}}\right)^{2} \approx
\begin{cases} 
1 & \text{for } f < f_{H_{\text{reh}}}, \\
\sqrt{\frac{\xi(f)-1}{\xi(f)+1}} & \text{for } f_{H_{\text{reh}}} < f < f_*, \\
\sqrt{\frac{\xi(f_*)-1}{\xi(f_*)+1}} & \text{for } f > f_*.
\end{cases}
\end{equation}
Let us note that the frequency dependence in Eq.(\ref{finred}) originates from the red-shifting of the tensor modes after they re-enter the Hubble horizon and propagate.

\begin{figure}[ht]
   	\includegraphics[height = 8cm,width=12cm]{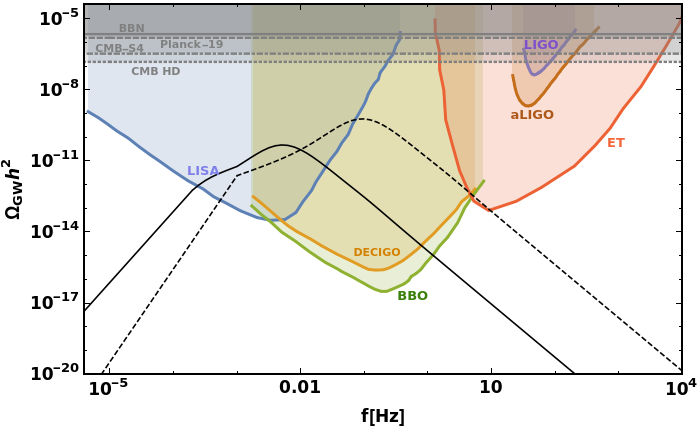}
    \caption{\it { The GW spectrum with EMD (Solid Black curve) and without EMD (Black Dashed Curve), along with the reaches of the future observations and the region already excluded. The parameter choice are: $\,\alpha=100,\,\beta/H_{*}=10,\, \kappa_{\text{col}}=0.999$, $f_{*}=10^{-3}$, $M_{\phi}=10^{12} \,\text{GeV},$
$M_{N_1}=10^{10}\, \text{GeV},\, \lambda=10^{-10} $ with $\Gamma_\phi /H_{*}\sim10^{-2}$.} }
	\label{GW_RDMD}
\end{figure}

\begin{figure}[ht]
   	\includegraphics[height = 8cm,width=12cm]{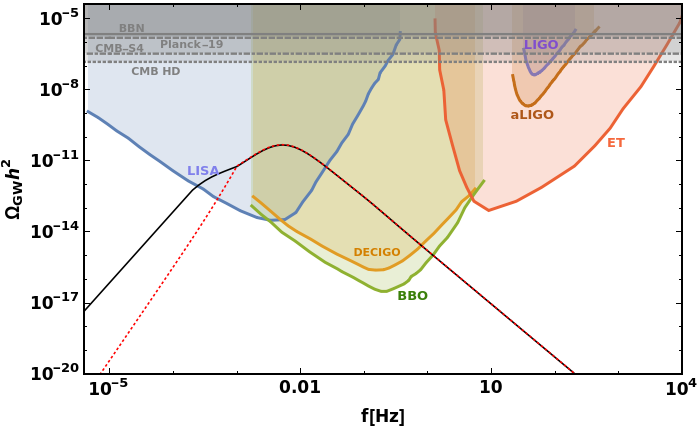}
    \caption{\it {The GW spectrum with EMD (Black solid curve) and with RD (Red dotted curve), along with the reaches of the future observations and the region already excluded. The parameter choices are: $\,\alpha=100,\,\beta/H_{*}=10 \, (\text{EMD}), 35\, (\text{RD}),\, \kappa_{\text{col}}=0.999$, $f_{*}=10^{-3}$, $M_{\phi}=10^{12} \,\text{GeV},$
$M_{N_1}=10^{10}\, \text{GeV},\, \lambda=10^{-10}\,(\text{EMD}),\,6.3\times 10^{-9}\,(\text{RD}) $ with $\Gamma_\phi /H_{*}=10^{-2}(\text{EMD})$.} }
	\label{GW_RDMD_2}
\end{figure}

We illustrate the GW spectrum in the presence of matter-domination and compare it with the standard radiation-domination in Fig.\ref{GW_RDMD}, along with the various experimental sensitivities of GW detectors. The black and black-dashed lines show the GW spectrum with and without EMD, respectively. In the presence of early matter domination, the signal suppresses and moves towards the lower frequency region, depending on the ratio $\Gamma_{\phi}/H_{*}<1$. The redshifting of the peak frequency in the case of MD is more, as with a low value of $\Gamma_{\phi}$ (required for a higher duration of matter domination), the peak frequency also decreases, which follows from Eq.(\ref{arehas}). Our parameter choice is given in the caption of the Fig.\ref{GW_RDMD}. For small values of the Yukawa coupling $\lambda$, the GW signal falls in mid-frequency ranges that are inspected by the LISA, DECIGO, and BBO. In Fig. \ref{GW_RDMD}, the regions above the horizontal lines labeled as BBN and Planck-2019, and the shaded region as LIGO are already excluded regions from available data.

In Fig.\ref{GW_RDMD_2}, we show another comparison of the GW spectrum with EMD (Black Solid) to the one in the standard RD scenario (Red Dotted). Here, the solid black curve is the same as in Fig.\ref{GW_RDMD}, while we set the parameters for the standard case (Red dashed curve) to well overlap the peak region. It is evident that in the EMD case, there is a change in the slope of the spectrum for lower frequencies, which is the key to distinguishing the cosmological history with EMD from the standard one.

\begin{figure}[ht]
   	\includegraphics[height = 8cm,width=12cm]{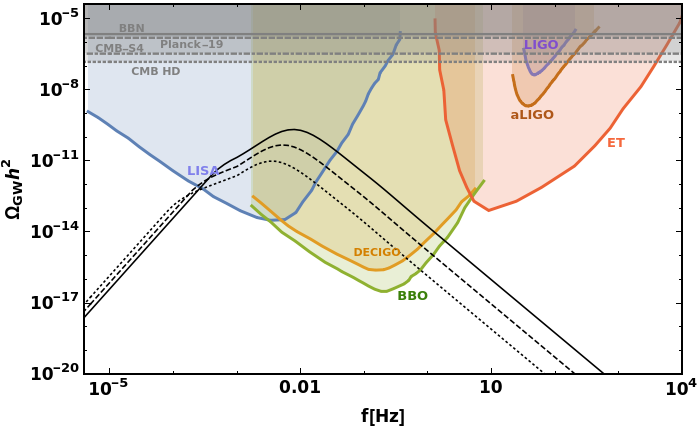}
    \caption{\it  The GW spectrum with EMD: The parameter choices are: $\,\alpha=100,\,\beta/H_{*}=10,\, \kappa_{\text{col}}=0.999,$ 
$\, f_{*}=10^{-3}, $
$M_{\phi}=10^{13} \,\text{GeV},\text{(solid)} ,$ 
$M_{\phi}=10^{12} \,\text{GeV}\,\text{(dashed)},$
$M_{\phi}=10^{11} \, \text{GeV} \,\text{(dotted)},$
$ M_{N_1}=10^{10}\, \text{GeV},\, \lambda=10^{-10}$, with $\Gamma_\phi/H_{*}= 10^{-1}\,\text{(solid)},\,10^{-2}\,\text{(dashed)},\,10^{-3}\,\text{(dotted)}$. }
	\label{GW_MD}
\end{figure}

In the radiation-dominated era, the spectrum below the peak is proportional to $f^3$, but it changes to $f^1$ during the intermediate matter-domination era \cite{Gouttenoire:2021jhk,Ellis:2020nnr}. The range of the frequency is determined by $\frac{\Gamma_{\phi}}{H_{*}}$ as we have assumed that the matter domination begins at $H_{*}$. In Fig.\ref{GW_MD}, we show the behavior of the GW spectrum for different choices of the ratio $\Gamma_\phi/H_{*}\sim 10^{-1},\,10^{-2},\,10^{-3}$. For decreasing values of the $\Gamma_\phi/H_{*}$ ratio, the overall abundance of the GW spectrum is suppressed due to modified red-shifting, whereas the plateau of the GW spectrum elongates at lower frequencies. The parameter choice is given in the figure caption. For the chosen $\lambda$, the GW signal may be detectable by LISA, whose detection covers the GW spectrum for lower and higher frequencies around the peak.

\section{Results: Leptogenesis and the Gravitational Waves}
\label{GW_BAU_prob}

As we discussed, the scalar field $\phi$ dominates the energy budget of the Universe. The produced right-handed neutrino from the $\phi$ decay sources the generation of the lepton asymmetry via its subsequent decay. This asymmetry is transferred to the observed baryon asymmetry of the Universe via the electroweak sphalerons. As we illustrated in the previous section, the duration of the MD of the scalar field related by its decay width, given by Eq.(\ref{dur_MD}), may leave imprints on the Gravitational Wave signals. Both scenarios are related by the decay width of the scalar field $\Gamma_\phi$. The observation of the GW spectrum may help to reconstruct the particle physics parameters like $M_{\rm \phi}$ and $\Gamma_{\phi}$.

\begin{figure}[ht]
  \includegraphics[height = 5.5cm,width=8cm]{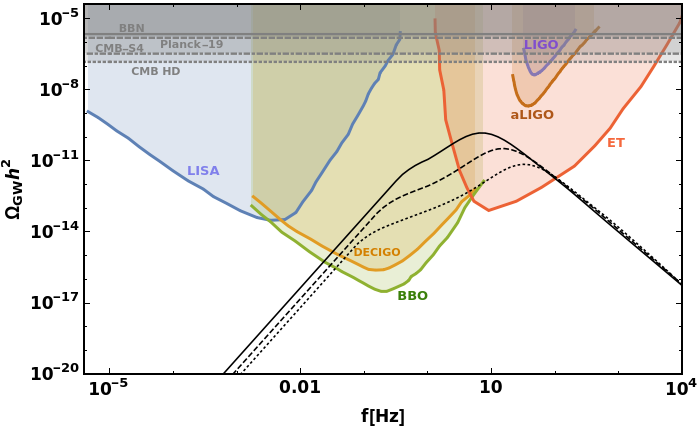}
  \includegraphics[height = 5.5cm,width=8cm]{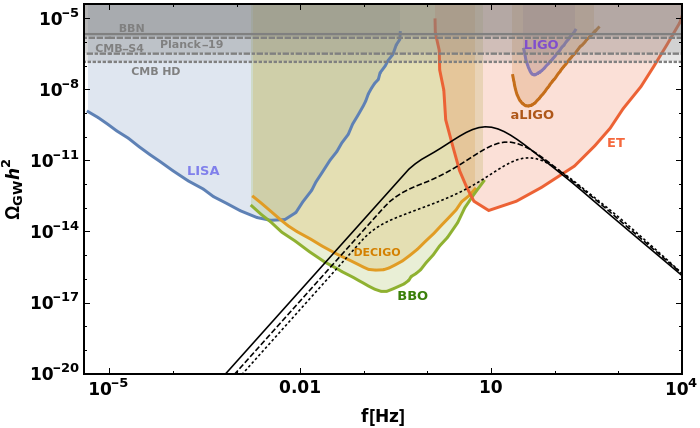}
  \includegraphics[height = 5.5cm,width=8cm]{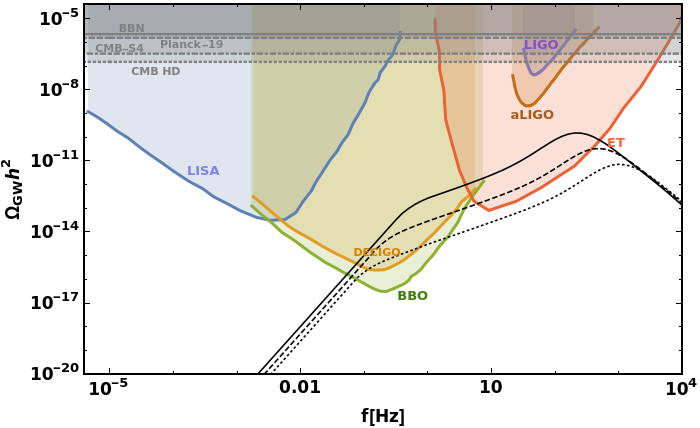}
  \includegraphics[height = 5.5cm,width=8cm]{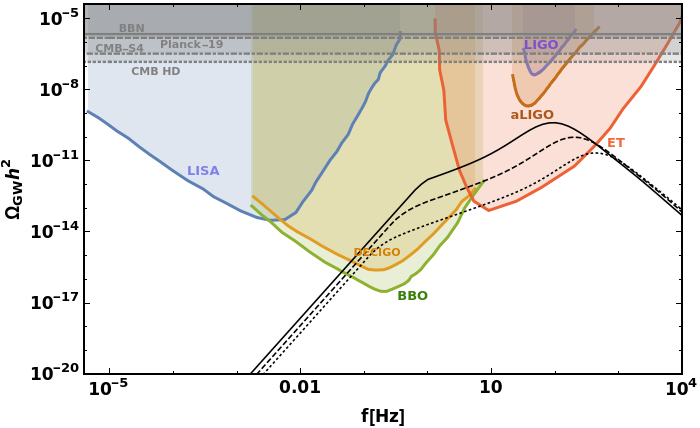}
    \caption {\it The GW spectrum, which is compatible with the BAU. The solid line corresponds to BPs given in Table \ref{table_BPs}. Solid, dashed and dotted lines correspond to $\Gamma_{\phi}/H_{*}\sim 10^{-1},\,10^{-2}, {\rm and}\,10^{-3}$ respectively.} \label{figBPs}
 \end{figure}

To analyze the parameter space of both successful Leptogenesis and the GW spectrum, we present two benchmark points (BPs) for each NH and IH case in Table \ref{table_BPs}. Corresponding to each BP, we show the GW spectrum plots in Fig.\,\ref{figBPs}. The first and second rows of the figure correspond to NH and IH cases, respectively. For both NH and IH cases, the peak regions of the GW spectrum can be covered by the ET, and the plateau below the peak may be covered by ET, DECIGO, and BBO. We observe that, compared to NH, the IH case GW spectrum falls in the high frequency region. For each BP we consider, the results for different choices of $\Gamma_{\phi}/H_*\sim 10^{-1}\,(\text{solid}),\,10^{-2}\,(\text{dashed}), {\rm and}\,10^{-3}\, (\text{dotted})$ are also shown.

\begin{table}[ht]
\begin{center}
\begin{tabular}{|c|c|c|c|c|c|}
\hline
  & NH BP 1 & NH BP 2 & IH BP 1 & IH BP 2\\
\hline
$M_{\phi}$ (GeV)  & $10^{14}$ & $5 \times 10^{13}$ & $10^{14}$& $ 5\times 10^{14}$ \\\hline
$M_N$ (GeV)    & $10^{13}$ & $10^{13}$ & $10^{13}$ & $10^{13}$ \\
\hline
$\lambda$  & $2.51\times 10^{-8}$ & $6.3\times 10^{-8}$ & $7.94\times 10^{-7}$ & $2.51\times 10^{-7}$ \\\hline
$a+i\,b$  & $1.35+ i \,1.57$ & $1.7+ i\, 1.57$ & $1.19+ i \,1.57$  & $0.94+i \,1.57$  \\ \hline
$Y_B$ & $8.7\times 10^{-11}$ & $8.7\times 10^{-11}$ & $8.7\times 10^{-11}$ & $8.7\times 10^{-11}$ \\ \hline
$T_{\text{RH}} $ (GeV) &  $2.88\times 10^{7}$  & $4.65\times 10^{7}$ & $9.13 \times 10^{8}$ & $6.65\times 10^{8}$  \\ \hline
$H_{*}$ (GeV) & $10^{-2}$ & $10^{-2}$ & 10 &  $1$ \\\hline
$f_{*} [Hz]$ & $1$ & $1$ & $1$ & 1 \\\hline
$\Gamma_{\Phi}/H_{*}$ & $0.12$ & $0.3$ & 0.12  & 0.63 \\\hline
\end{tabular}
\end{center}
\caption{\it Benchmark Points to study the matter domination effects in the GW spectrum that also satisfies the BAU (both for normal and inverted hierarchy cases).}
\label{table_BPs}
\end{table}

\subsection{GW Detection prospects with interferometers}
\label{subsec:detection_pros}

The strong first-order phase transition GW signals, as we discussed earlier, can be detected in the next generation of GW missions. There are a plethora of upcoming GW observations, with several of them, LISA and ET, scheduled to be operational in the 2030s. With such improved prospects of GW detection, one may broadly classify the GW experiments into the following categories:
\begin{enumerate}
    \item \textbf{Ground based GW interferometer missions:} \textit{Laser Interferometer Gravitational-wave Observatory} (LIGO)~\cite{LIGOScientific:2016aoc,LIGOScientific:2016sjg,LIGOScientific:2017bnn,LIGOScientific:2017vox,LIGOScientific:2017ycc,LIGOScientific:2017vwq}, \textit{Advanced} LIGO (a-LIGO)~\cite{LIGOScientific:2014pky,LIGOScientific:2019lzm},  \textit{Einstein Telescope} (ET)~\cite{Punturo_2010,Hild:2010id}, \textit{Cosmic Explorer} (CE)~\cite{Reitze:2019iox}.
    \item \textbf{Space based GW interferometer missions: }$\mu$-ARES~\cite{Sesana:2019vho}, \textit{Laser Interferometer Space Antenna} (LISA)~\cite{amaroseoane2017laser,Baker:2019nia}, \textit{Big-Bang Observer} (BBO)~\cite{Corbin:2005ny,Harry_2006}, \textit{Deci-Hertz Interferometer Gravitational-wave Observatory} (DECIGO)~\cite{Yagi:2011yu}, \textit{Upgraded} DECIGO (U-DECIGO)~\cite{Seto:2001qf,Kawamura_2006,Yagi:2011wg}.
    \item \textbf{Pulsar Timing Arrays (PTA):} \textit{European Pulsar Timing Array} (EPTA)~\cite{Kramer:2013kea,Lentati:2015qwp,Babak:2015lua}, \textit{Square Kilometer Array} (SKA)~\cite{Janssen:2014dka,Weltman:2018zrl,Carilli:2004nx}, \textit{North American Nanohertz Observatory for Gravitational Waves} (NANOGrav)~\cite{McLaughlin:2013ira,NANOGRAV:2018hou,Aggarwal:2018mgp,Brazier:2019mmu,NANOGrav:2020bcs}.
\end{enumerate} 

In the first set of Figs. \ref{GW_RDMD}-\ref{GW_MD}, we show the power-law integrated (PLI) sensitivity curves as discussed in~\cite{Thrane:2013oya} for several forthcoming GW experiments. PLANCK-19 and LIGO denote the present observational constraints. The rest of the color-shaded regions depict the future sensitivity prospects in each of the GW missions mentioned above. These power law sensitivity curves are drawn based on the following assumption: the expected GW spectrum from first-order phase transition can be represented in a power-law form, that is, $\Omega_{\rm GW} \sim f^b$, where $b$ is the spectral index of the frequency slope. Usually, the shaded region that falls inside such a PLI curve accounts for the range of parameters where such a power-law model GW signal will be detected with quite a high signal-to-noise ratio (SNR). Although these power law curves give us a useful way to represent phenomenological predictions, there are other methods to depict such a signal, for instance, peaked-integrated sensitivity curves (PISCs), which in certain phase transition scenarios may fare better \cite{Schmitz:2020syl}. Our assumption of a power-law in the context of this paper is justified in the sense due to the separate period of radiation and early matter domination, such regions have distinct powers of the frequency slopes, unlike the standard cosmological first-order phase transition, where there is only one slope exhibiting only a characteristic peak in the GW signal.  Frankly, a reliable detection is only ensured for signals integrated over the PLI framework; however, such an analysis is beyond the scope of the current paper and can be taken up in a future analysis.


 \medskip

\section{Conclusions and Discussion} \label{conclusion}

We considered a non-thermal leptogenesis scenario with a scalar field $\phi$, which not only dominates the Universe’s energy density in the early Universe but also couples to right-handed neutrinos (RHNs). As $\phi$ decays into RHNs, their subsequent out-of-equilibrium decay produces the lepton asymmetry for the successful baryogenesis. Simultaneously, this $\phi$ realizes a matter domination era, which leaves imprints on the gravitational wave (GW) spectrum from a strong first-order phase transition. We studied the effects of $\Gamma_\phi$, which played the key role in connecting the Leptogenesis with the GW spectrum, and showed that a lower $\Gamma_{\phi} / H_*$ ratio leads to stronger suppression in the GW amplitude and prolonged matter domination due to modified redshift history. If the characteristic plateau-like shapes of the GW spectrum shown in our results are indeed observed, one may reconstruct the BSM model parameters like masses and decay widths. However, with only the knowledge of the GW spectrum, distinguishing between the $\phi$-domination or any other forms of early matter domination may prove to be quite a difficult task. Then, one may look towards additional sources of information, and achieving the correct BAU may help to break such degeneracies.
We presented two benchmark points for each of the normal and inverted neutrino mass hierarchies, showing the common regions of parameter space that simultaneously yield successful leptogenesis and observable GW signals. The predicted spectra fall in the high frequency region measurable by the sensitivity of GW detectors such as ET\cite{Punturo:2010zz}, DECIGO\cite{Seto:2001qf, Kudoh:2005as, Kawamura_2006, Nakayama:2009ce, Yagi:2011wg, Kawamura:2020pcg}, and BBO \cite{Crowder:2005nr, Corbin:2005ny, Harry_2006}. In standard Type-I seesaw Leptogenesis models, the mass scale of Leptogenesis is very high and challenging to probe in a laboratory; therefore, measurements of such GW signals can be an alternative pathway to test for Leptogenesis. We would like to end by emphasizing the fact that testing the scale of new physics, if they are of high energy, like that of the scale of leptogenesis, is very challenging to test in laboratory experiments. Therefore, primordial gravitational waves, such as those from cosmological first-order phase transitions, can be quite useful and provide us with an alternative pathway toward the search for new physics in the future.

\vspace{1cm}
\medskip

\newpage
\section*{Acknowledgment}
The authors thank Marek Lewicki for useful discussions.  
KM acknowledges the financial support provided by the Indian Association for the Cultivation of Science (IACS), Kolkata. NN acknowledges the financial support provided by the IACS, Kolkata, where part of the work has done during RA-I position. The work of N.O. is supported in part by the United States Department of Energy
Grant Nos. DE-SC0012447 and DE-SC0023713.
\bibliographystyle{JHEP}
\bibliography{ref1}
\end{document}